\documentclass[]{emulateapj}

\shorttitle{Origin of M-Z Relation} 
\shortauthors{Brooks et al.}

\begin{document}

\title{The Origin and Evolution of the Mass-Metallicity Relationship for Galaxies: \\Results from Cosmological N-Body Simulations}

\author{A.\,M. Brooks\altaffilmark{1,2},
        F. Governato\altaffilmark{1},
        C.\,M. Booth\altaffilmark{3},
        B. Willman\altaffilmark{4,5},
        J.\,P. Gardner\altaffilmark{6},
        J. Wadsley\altaffilmark{7}, 
        G. Stinson\altaffilmark{1},
        T. Quinn\altaffilmark{1} 
}
\altaffiltext{1}{Astronomy Department, University of Washington, Box 351580, Seattle, WA, 98195-1580}
\altaffiltext{2}{e-mail address: abrooks@astro.washington.edu }
\altaffiltext{3}{Department of Physics, Institute for Computational Cosmology, University of Durham, South Road, Durham, DH1 3LE, UK}
\altaffiltext{4,}{Harvard-Smithsonian Center for Astrophysics, 60 Garden Street, Cambridge, MA, 02138}
\altaffiltext{5}{Clay Fellow}
\altaffiltext{6}{Department of Physics \& Astronomy, University of Pittsburgh, 100 Allen Hall, 3941 O'Hara Street, Pittsburgh, PA, 15260}
\altaffiltext{7}{Department of Physics and Astronomy, McMaster University, Hamilton, Ontario, L88 4M1, Canada}


\begin{abstract} 

We examine the origin and evolution of the mass-metallicity relationship
(MZR, {\it M$_*$-Z}) for galaxies using high resolution cosmological SPH 
+ N-Body simulations that include a physically motivated description of 
supernovae feedback and subsequent metal enrichment.  We discriminate between 
two sources that may contribute to the origin of the MZR: 1) metal and baryon 
loss due to gas outflow, or 2) inefficient star formation at the lowest galaxy 
masses.  Our simulated galaxies reproduce the observed MZR in shape and 
normalization both at z=0 and z=2.  We find that baryon loss occurs due to UV 
heating before star formation turns on in galaxies with {\it M$_{bar}$} $<$ 
10$^8$ M$_{\odot}$, but that some gas loss due to supernovae induced winds is 
required to subsequently reproduce the low effective chemical yield observed 
in low mass galaxies.  Despite this, we show that low star formation 
efficiencies, regulated by supernovae feedback, are primarily responsible for 
the lower metallicities of low mass galaxies and the overall {\it M$_*$-Z} 
trend.  We find that the shape of the MZR is relatively constant with redshift, 
but that its normalization increases with time.  Simulations with no energy 
feedback from supernovae overproduce metals at low galaxy masses by rapidly 
transforming a large fraction of their gas into stars.  Despite the fact that 
our low mass galaxies have lost a majority of their baryons, they are still 
the most gas rich objects in our simulations due to their low star formation 
efficiencies.

\end{abstract}


\keywords{galaxies: evolution --- galaxies: formation --- methods: N-Body simulations}


\section{Introduction}
\label{intro}

The observed trend that the metallicity of nearby galaxies decreases 
with galaxy mass has been well established in the literature 
\citep[e.g.,][]{T04,Lee06}.  This trend has also recently been
observed at high redshifts \citep[][hereafter S05 and E06, 
respectively]{Savaglio05,Erb06}.  The gas metallicity of a galaxy 
depends on reprocessed material from its stars and subsequent inflow 
and outflow of gas.  The mass-metallicity relationship (MZR) thus 
provides insight into two physical mechanisms important to the 
evolution of galaxies: star formation efficiency and gas 
inflow/outflow.

In a ``closed box'' system (no inflow or outflow of material),
metallicity {\it Z} is given by
\begin{equation}
Z = y\, {\rm ln} f_{gas}^{\rm -1},  
\label{eq:1}
\end{equation}
where {\it y} is the nucleosynthetic metal yield produced by stars 
and {\it f$_{gas}$} is the gas mass fraction (M$_{gas}$/[M$_{gas}$ + 
M$_{stars}$]) \citep[e.g.,][]{Tinsley}.
In general, low mass galaxies are observed to have higher gas 
fractions than high mass galaxies \citep[e.g.,][]{Geha06, West05}. 
This observation can easily explain the origin of the
{\it M$_*$-Z} relation (MZR) in the context of the closed box model.  
In this scenario, low mass galaxies are gas rich because they are 
inefficient at turning gas into stars, likely due to their low 
surface densities \citep{Kennicutt98, MK01, Verde02, JD04}.  

Alternatively, SNe feedback might expel gas in galactic 
winds.  In this scenario the loss of metals becomes progressively 
more important at lower masses due to shallower halo potential wells 
\citep[for a review see][]{vcb05} and leads to the observed 
{\it M$_*$-Z} trend. 
However, the importance of supernova feedback as a function of galaxy 
mass has been investigated by various groups, with results that are 
still debated \citep[][E06]{vcb05, Geha06, JD06}. 

Recently, studies have investigated how the MZR evolves with cosmic 
time \citep[S05, E06,][]{Dave06, deRossi06, kobayashi06, tkg06}.  Fully 
cosmological numerical experiments are one of the best tools to infer the 
relative importance of star formation (SF) and feedback processes in a 
realistic setting \citep[e.g.,][hereafter G06]{Okamoto05, Robertson06, 
Governato06}.  In this Letter, we investigate the origin and evolution of 
the MZR in a concordance $\Lambda$CDM cosmological setting with N-body + SPH 
simulations which include a physically motivated treatment of SF and SN 
feedback.  These simulations allow us to study the formation history of 
galaxies over range of masses, and thus quantify the importance of feedback 
and blowout versus low SF efficiencies over their lifetimes. 

\section{The Simulations}
\label{sims}

The simulations used in this study are the culmination of an effort to 
create realistic disk galaxies.  As discussed in G06, at z=0 disk 
galaxies in our simulations fall on the Tully-Fisher (TF) and baryonic TF 
relations \citep{G97, McGaugh05} and the age-mass relation \citep{Mac04}.  
This is and important improvement over previous works, and is made possible 
due to the inclusion of a simple but physically 
motivated recipe \citep[full details in][]{Stinson06} to describe SF and 
the effects of subsequent SNe feedback.  Our adopted SN feedback and 
cosmic UV background \citep{HM96} also drastically reduce the number of 
galaxy satellites containing a significant stellar population, making 
many of them ``dark'' \citep{Quinn96, Moore99}.  Thus, the SN and 
SF efficiency parameters were adopted as in G06, with an IMF from 
\citet{Kroupa}, with no additional tuning of the parameters to affect metals. 
Metal enrichment from both SN Ia\&II is followed based on the prescription 
of \citet{Raiteri}, who adopt yields from \citet{Thielemann} for SN\,Ia, and 
\citet{WW93} for SN\,II.

We selected four ``field'' regions (d$\rho$/$\rho$ $\sim$ 0.1)
from a low resolution, dark matter (DM) simulation run in GASOLINE 
\citep{Wadsley} using a concordance, flat, $\Lambda$-dominated cosmology:
$\Omega_0=0.3$, $\Lambda$=0.7, $h=0.7$, $\sigma_8=0.9$, shape
parameter $\Gamma=0.21$, and $\Omega_{b}=0.039$ \citep{Perl97, E02}.
These field regions were resimulated at higher resolution using the 
volume renormalization technique \citep{KW93}, which allows us to 
resolve fine structure while capturing the effect of large scale torques.  
The field regions are centered around galaxies with peak rotation 
velocity V$_{rot}$ $\sim$ 150 and 275 km/s, with each high resolution 
region having a comoving volume of  $\sim$100--1000 Mpc$^3$.  Two of 
the main galaxies are described in detail in G06, but for this 
study they have been rerun at higher resolution.  At z=0, these runs 
yield a sample of 31 high resolution disk and dwarf galaxies over a range 
of peak velocities 30 km/s $<$ V$_{rot}$ $<$ 275 km/s, or total halo 
masses from 3.4$\times$10$^9$ M$_{\odot}$ to 1.1$\times$10$^{12}$ 
M$_{\odot}$.  Seven of these 31 galaxies are satellites of the primaries.  
Galaxies and their parent halos were identified using 
AHF\footnote{{\bf{A}}MIGA's {\bf{H}}alo {\bf{F}}inder, available for 
download at http://www.aip.de/People/aknebe/AMIGA} \citep{Knebe, GK04}. 
Our highest (lowest) resolution runs have particle masses of 
7.6$\times$10$^5$ M$_{\odot}$ (4.9$\times$10$^6$ M$_{\odot}$), 
1.3$\times$10$^5$ M$_{\odot}$ (8.5$\times$10$^5$ M$_{\odot}$), and 
3.8$\times$10$^4$ M$_{\odot}$ (2.5$\times$10$^5$ M$_{\odot}$) for DM, 
gas, and stars, respectively, and a force resolution 
of 0.15-0.3 kpc.

\section{The Origin of the MZR} 
\label{origin}

\begin{figure}
\plotone{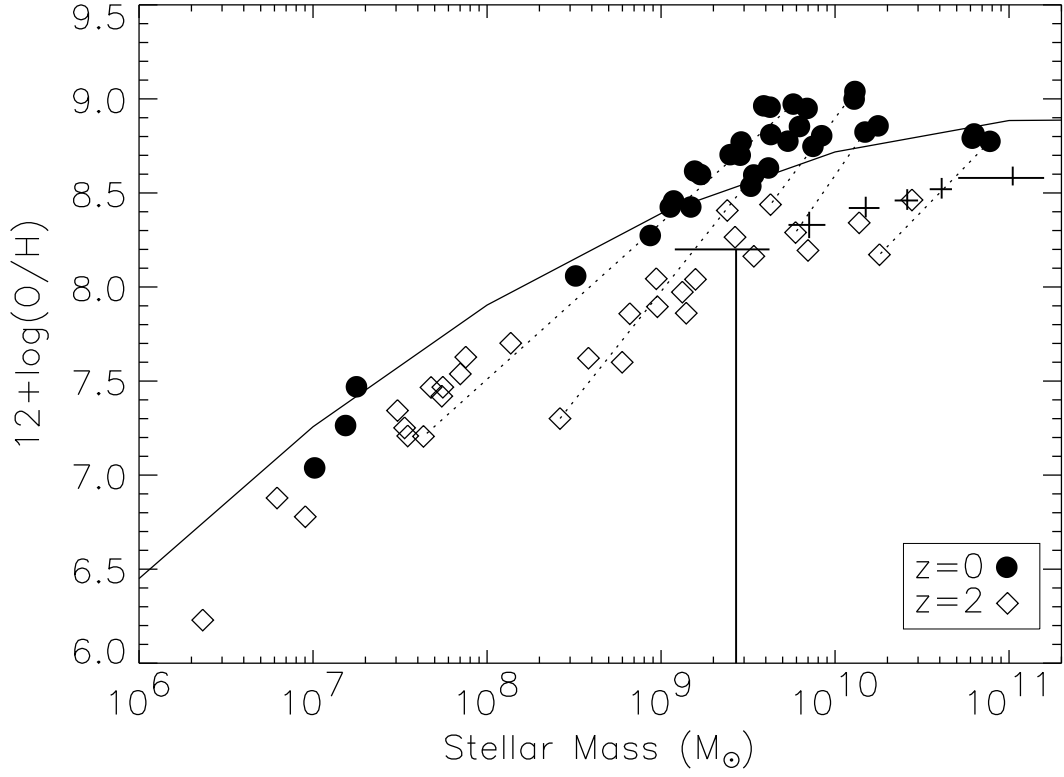}
\caption{The mass-metallicity relation of our simulated 
galaxies at z=0 (solid points) and z=2 (open diamonds).  The solid 
curved line is the observational fit to $>$53,000 galaxies in SDSS 
from T04, shifted down by -0.26 dex as found by E06.  Error bars show 
the observational mass-metallicity relation at z=2 from E06.  Dotted 
lines connect some of the z=0 galaxies to their progenitors at z=2, 
showing how galaxies evolve in the M$_*$-Z plane with time.  }
\label{fig1}
\end{figure}

The MZR for 31 high resolution galaxies in 
our simulations is shown in Fig.~\ref{fig1}, at both z=0 and z=2.  The  
solid curved line is the empirical fit for $>$ 53,000 galaxies in SDSS 
from \citet[][hereafter T04]{T04}.  It has been shifted down by 0.26 dex 
as found by E06 to correct for the fact that the method used by T04 is 
known to systematically find O abundances up to 0.3 dex larger than other 
methods \citep[T04, S05, E06][]{Lee06}.  The galaxies in our sample at 
z=0 match the empirical fit of T04 extremely well, reproducing
the observed trend over almost 4 orders of magnitude in stellar 
mass.  We note that our lowest mass galaxies, however, have a lower O 
abundance than that found for dwarf irregular galaxies by \citet{Lee06}.  

Fig.~\ref{fig1} also shows that the MZR for our simulated galaxies at 
z=2 (open diamonds) is in excellent agreement with high redshift data.  
The error bars overplotted on Fig.~\ref{fig1} display the observational 
z=2 results from E06.  We have identified in the simulations the z=0 
galaxies corresponding to those plotted at z=2.  The dotted lines in 
Fig.~\ref{fig1} connect a few of these galaxies, which are representative 
of how all the galaxies have evolved in the {\it M$_*$-Z} plane with time.  

To make a meaningful comparison with the T04 sample, Fig.~\ref{fig1} uses 
only cool gas particles (with temperature $<$ 40,000 K) within the radius 
of each galaxy that encloses 2/3 of the stellar mass.  The median 
``observed'' diameter for our galaxies is 5 kpc, similar to that found 
by T04.  At z=2 we included all cold gas in each galaxy.  
Even if we include all cool gas at z=0, the {\it M$_*$-Z} trend is 
unchanged up to 10$^{10}$ M$_{\odot}$, showing that the evolution seen 
in Fig.~\ref{fig1} is not an effect of our measurements. 

A few runs were repeated at lower resolutions (1/2 and 1/20 of the
mass resolution).  Low resolution runs undergo less SF, 
thus decreasing both M$_*$ and the total amount of O produced.  
Our results from each run begin to significantly diverge from each other 
($\Delta$M$_*$ $>$ 20\%, $\Delta$12+log(O/H) $>$ 0.2 dex) if the number 
of DM particles in a halo drops below a few thousand and/or mass and 
force resolution degrade too much.  Note that this low resolution effect 
can lead to an artificially steeper {\it M$_*$-Z} trend, making it essential 
that results converge.  Informed by these resolution tests, we have only 
included galaxies with more than 3500 DM particles in our sample (at 
high and low redshift), ensuring that their stellar and metal content 
have converged.

Having shown that our simulations reproduce the observed MZR at both 
z=0 and z=2, we will next examine the origin of this relation.  We focus 
on the importance of preferential metal loss versus low SF efficiency 
at low masses in producing the MZR.

\subsection{The Role of Mass Loss}
\label{massloss}

\begin{figure}
\plotone{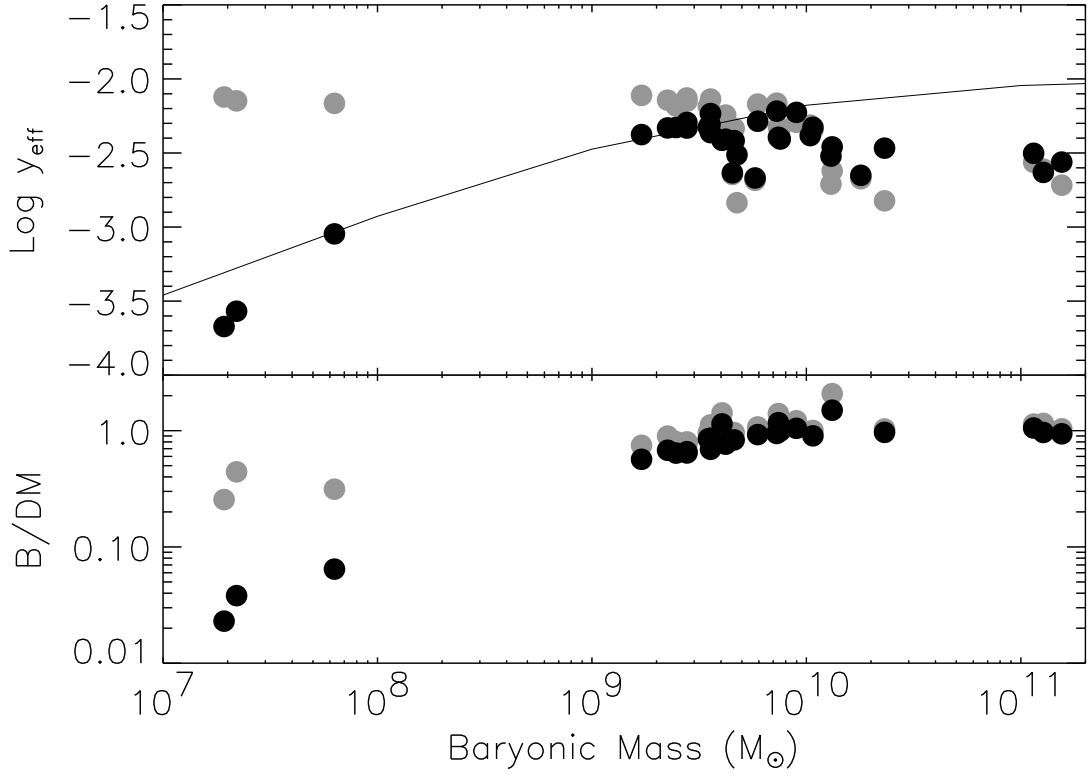}
 \caption{ {\it Top Panel}: Effective yield vs total
 baryonic mass for our galaxies at z=0 (black circles).  
 The solid curved line is the observational fit to $>$53,000 galaxies 
 from T04.  
 {\it Bottom Panel}: Baryonic to dark matter mass ratio (relative to 
 the cosmic abundance) vs total baryonic mass.  Satellite galaxies 
 excluded.  In both panels, grey circles show what the value would be 
 if no gas loss from the galaxies had occurred. }
\label{fig2} 
\end{figure}

An observable quantity commonly used to investigate mass loss from 
galaxies is the effective yield, {\it y$_{eff}$}.  From Eq.~\ref{eq:1}, we 
can define the effective yield,
\begin{equation}
y_{eff} = Z/({\rm ln} f_{gas}^{\rm -1}).
\label{eq:2}
\end{equation}
In a closed box system, {\it y$_{eff}$} is equal to the true stellar
nucleosynthetic yield, but it can be inferred from 
Eq.~\ref{eq:2} that {\it y$_{eff}$} may deviate from the true yield due to 
inflow or outflow of gas.  The top panel of Fig.~\ref{fig2} shows 
{\it y$_{eff}$} for our galaxies, derived again using inner cold gas 
abundances, versus the total baryonic mass of our galaxies at z=0 (black 
circles).  The effective yields reproduce the observed trend from T04 (solid 
line) over most of our mass range.  Our two lowest mass galaxies have 
{\it y$_{eff}$} below that found by \citet{Lee06}, likely because 
these two have undergone additional tidal stripping of gas. 

The black circles in the bottom panel of Fig.~\ref{fig2} show the 
baryonic to dark mass ratio (B/DM) of our galaxies, normalized by the 
cosmic ratio (the seven satellite galaxies have been excluded from this 
panel).  In our simulations, large deviations below B/DM = 1 can only 
be produced by mass loss.  Our lowest mass galaxies have lost up to 98\% of 
their baryonic mass.  

Using the halos defined by AHF (our halo finder), we identify a given halo 
and its gas particles back to z=3.  Once all the gas particles that have ever 
belonged to a galaxy back to z=3 are identified, those particles that are no 
longer within the virial radius at z=0 are considered to be gas ``lost'' from 
the galaxy.  The grey points in both panels of Fig.~\ref{fig2} represent 
this information, for all gas (remaining + lost). 

When we include all of the gas lost since z = 3, we find that our low mass 
galaxies still have B/DM that fall significantly below the cosmic ratio.  
The bottom panel of Fig.~\ref{fig2} show that our lowest mass galaxies have 
lost 50 - 90\% of their baryons prior to z = 3.  We verify that a 
UV only simulation (SNe feedback turned off) removes up to 80\% of baryons 
from our lowest mass galaxies, but that these galaxies evolve as closed boxes 
once SF commences.  We thus find that SNe induced mass loss can account 
for up to 20\% of baryon loss from galaxies with 
M$_{bar}$ $<$ 1$\times$10$^8$ M$_{\odot}$.  Additional gas loss can occur 
via tidal stripping.  

Thus, Fig.~\ref{fig2} clearly demonstrates that gas has been preferentially 
lost from our lowest mass galaxies, and that this gas loss is responsible for 
the observed {\it y$_{eff}$}-M$_{bar}$ trend.  However, when all gas 
(remaining + lost) is used to rederive the MZR (Fig.~\ref{fig1}), the 
observed trend that lower mass galaxies have lower metallicities still 
exists, and it cannot be mass loss alone that causes the MZR. 

\subsection{Star Formation Efficiencies}
\label{sfe} 

As explained above, the mass loss experienced by our lowest mass galaxies 
cannot account for the overall {\it M$_*$-Z} trend seen in Fig.~\ref{fig1}.  
We now investigate whether SF efficiencies can explain the origin of 
this trend.  
If lower mass galaxies convert gas into stars less efficiently than 
higher mass galaxies, then they will require a much longer period of 
time to enrich the remaining gas to the same metallicity as higher mass 
galaxies.  This effect can lead to the observed MZR, 
with or without mass loss as an additional contributor.  

We verified that all of our galaxies follow the Schmidt-Kennicutt law for 
SF \citep{Kennicutt98, MK01}.  Our galaxies span a factor of 250 in gas 
surface densities, yielding a SFR that varies by over three decades from 
our lowest to highest mass galaxies.  Our lowest mass galaxies are thus 
incredibly inefficient at SF compared to our high mass galaxies.  The 
consumption rate (SFR/M$_{coldgas}$) of our lowest mass galaxies is about 
two decades lower than our highest mass galaxies.  That is, while our 
highest mass galaxies will use up all their gas in 1 Gyr or less, our 
lowest mass galaxies will take up to 60 Gyr if they continue SF at 
their current rate.  

\begin{figure}
\plotone{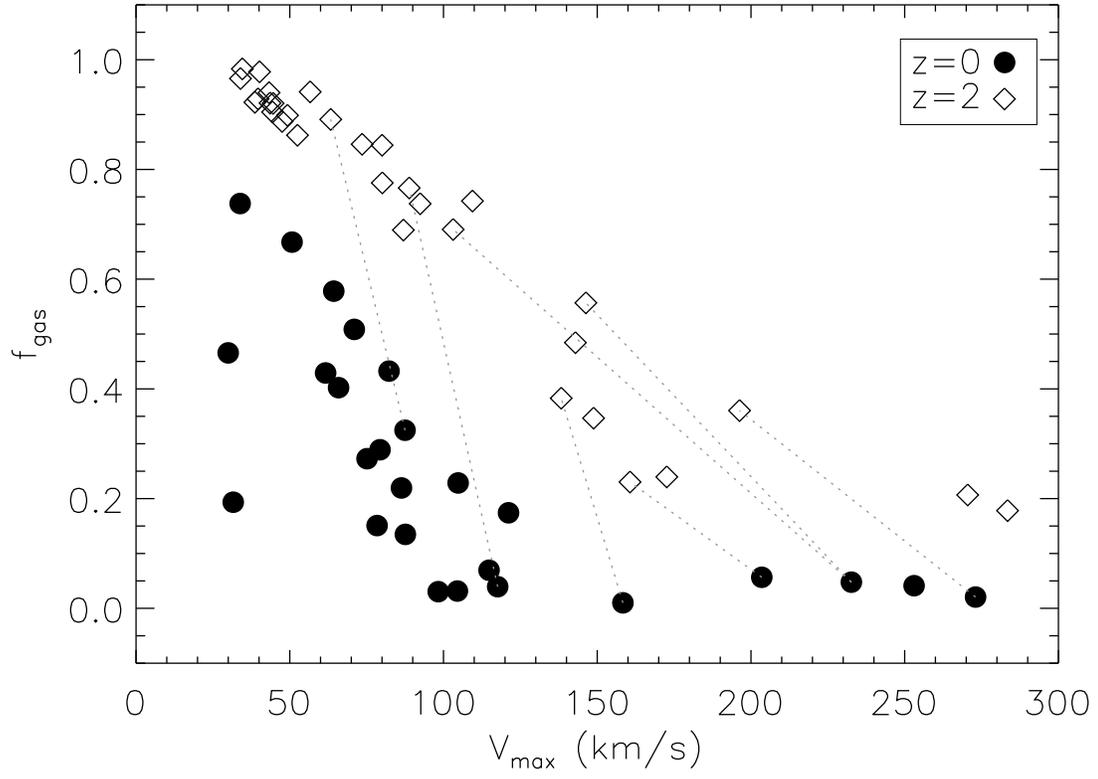}
\caption{The gas mass fraction vs V$_{rot}$ for our 
galaxies at z=0 and z=2 (satellite galaxies excluded).  Faint dotted 
lines connect a few of the galaxies at z=2 to their evolved counterpart 
at z=0. } 
\label{fig3}
\end{figure}

Fig.~\ref{fig3} shows the gas mass fraction, {\it f$_{gas}$}, versus 
rotational velocity for our galaxies (satellite galaxies excluded).  As 
found by recent observations of isolated galaxies \citep[e.g,][]
{West05, Geha06}, lower mass galaxies have increasing gas fractions, 
implying that lower mass galaxies have been less efficient at making 
stars.  Note that while the lowest mass galaxies in our sample have 
lost more than 90\% of their baryons (Fig.~\ref{fig2}), they remain 
gas rich due to their low SF efficiencies (with the exception of two 
that have been tidally stripped of much of their gas).    

We reran our simulations without SNe feedback.  In our models, 
without this feedback the MZR is flat because low mass galaxies
experience unrealistically high SF rates, resulting in an
overproduction of metals.  This difference with the runs that do
include SNe demonstrates that energy injection by SNe into the ISM
regulates SF efficiency, independent of whether SNe also
lead to mass loss.  SNe feedback thus indirectly affects the shape of
the MZR by modulating the gas surface densities, and hence
the SFR, with galaxy mass.

\section{The Evolution of the MZR} 
\label{evolution} 

Fig.~\ref{fig3} shows that our simulated galaxies at z=2 are less 
evolved than their z=0 counterparts, with z=2 galaxies being much more 
gas rich at a given rotational velocity than at z=0.  By z=0, they 
have consumed more of their gas, increasing both their stellar mass and 
O abundance; hence the evolution of the MZR as shown in Fig.~\ref{fig1}.  

Both S05 and E06 examined the MZR at z$\sim$0.7 and z$\sim$2, respectively, 
and drew conclusions similar to ours.  As pointed out by S05 and 
\citet{Pettini06}, at a given metallicity the high z data are about an 
order of magnitude more massive than the z=0 trend (Fig.~\ref{fig1}).  This 
suggests that more massive galaxies enrich at a faster rate than lower mass 
galaxies.  This is supported by our results in section~\ref{sfe}.  Both S05 
and E06 conclude that the evolution of the MZR is best described by a model 
in which low mass galaxies are inefficient at SF relative to high mass galaxies 
for a similar formation epoch.  Thus the period of SF is lengthened in lower 
mass galaxies and they remain more gas rich at a given redshift.  

We find that a shift downward by 0.3 dex in $\Delta$log(O/H) from fit shown 
in Fig.~\ref{fig1} is an excellent fit to our z=2 data, as was also found 
by E06.  A shift in $\Delta$log(M$_*$) is much too steep to match our 
simulations.  Thus, we find that the slope of the MZR remains relatively 
constant with redshift, but the normalization increases with time.

\section{Conclusions} 

We have presented results from high resolution simulations of field
galaxies in a $\Lambda$CDM context. The simulations are carried to
z=0 and include SF and a physically motivated description
of the effects of SNe feedback and metal enrichment. We have shown 
in previous work (G06) that this approach leads to disk galaxies that
fall on the TF and baryonic TF relations and reproduce the age-mass 
relation of disk galaxies and abundance of galaxy satellites.  Here we 
present a natural extension of this analysis, showing that the simulated 
galaxies are also in excellent agreement with the observed MZR both at 
z=0 and z=2.   

Simulated galaxies match both the slope and normalization of the MZR  
over four orders of magnitude in stellar mass. We find that 
the slope of the MZR is roughly constant with redshift, but that its 
normalization increases with time.  Such evolution agrees with the high-z 
observations of E06, and with early theoretical predictions \citep{Dave06, 
deRossi06}.  Furthermore, the mass scales of the turnovers in our 
simulated {\it Z}, {\it y$_{eff}$}, B/DM, and {\it f$_{gas}$} versus
mass relations are in good agreement with a range of previous 
observational and theoretical studies that find critical transitions 
at $\sim$120 km/s \citep[e.g.,][]{JD04, Keres, DB06}. 

Galaxies with M$_{bar}$ $<$ 10$^{8}$ M$_{\odot}$ (V$_{rot}$$\sim$50
km/s), lose a significant amount of their baryons due to combined 
heating from the cosmic UV field and SN feedback.  Baryon loss due to
SN feedback results in a 
{\it y$_{eff}$}-M$_{bar}$ trend in agreement with the results from 
the SDSS (T04).  However, including gas that has been lost from 
galaxies does not alter the O abundances significantly, and does not 
change the overall shape of the MZR, confirming that 
the lower O abundances in our low mass galaxies are primarily due to 
their low SF efficiencies rather than directly to blowouts.  Despite 
strong baryon loss from their halos, we find that our cold gas 
fractions at low galaxy masses are in qualitative agreement with recent 
observational samples \citep{Geha06}.

SN feedback plays a crucial role in lowering the SF efficiency in low 
mass galaxies and originating the turnover in the MZR at the observed 
scales.  Without energy injection from SNe to regulate SF, gas that 
remains in galaxies rapidly cools, forms stars, and increases the O 
abundance, even by z=2 (see G06 for additional shortcomings of the no 
feedback case). With feedback from SNe turned off, small 
galaxies produce too many metals too early at the expense of their 
cold gas content, producing {\it M$_*$-Z} and {\it y$_{eff}$}-M$_{bar}$ 
relations too flat compared to observations, in agreement with some 
early theoretical predictions \citep{deRossi06, JD06}.  SN feedback 
prevents low mass galaxies from enriching their gas to a high 
metallicity at all redshifts.  In our runs, star formation in small 
field galaxies occurs over extended periods of time until the present, 
in agreement with observations \citep{Dolphin05}.


\acknowledgments

We would like to thank C. Brook, H. Lee, E. Skillman, K. Venn, 
J. Dalcanton, A. Kravtsov, and A. West for helpful conversations 
during this project.  Support for this work was provided by the 
Spitzer Space Telescope Theoretical Research Program.  FG 
acknowledges support from NSF grant AST-0098557.  Simulations were 
run at the Pittsburgh Supercomputing Center, SDSC, and Cineca.   
AB, GS, and TQ were supported by NSF ITR grant PHY-0205413.  


\clearpage 

\clearpage

\clearpage

\end{document}